\documentclass[aps,twocolumn,prd, superscriptaddress,nofootinbib]{revtex4-2}
\pdfoutput=1

\usepackage{amsmath,amssymb}
\usepackage{graphicx}
\usepackage{xcolor}
\usepackage[colorlinks=true,citecolor=blue,linkcolor=blue,urlcolor=blue]{hyperref}
\usepackage{placeins}
\usepackage{float}

\newcommand{\Pcal}{\mathcal P}
\newcommand{\Ccal}{\mathcal C}
\newcommand{\eps}{\epsilon}
\newcommand{\Jcp}{J_{\rm CP}^{\ell}}
\newcommand{\Jmax}{J_{\max}}
\newcommand{\dCP}{\delta_{\rm CP}^{\ell}}
\newcommand{\VRl}{V_R^\ell}
\newcommand{\VL}{V_L}
\newcommand{\diag}{\mathrm{diag}}
\newcommand{\Tr}{\mathrm{Tr}}
\newcommand{\OO}{\mathcal O}
\newcommand{\CP}{\mathrm{CP}}
\newcommand{\CPell}{\mathrm{CP}_\ell}

\begin{document}

\title{Strong CP and the PMNS Phase in Dirac Left-Right Symmetry}

\author{Vladimir Tello}
\affiliation{University of Split, FESB, Croatia}
\date{26 August 2026}

\begin{abstract}

Left--right symmetric theories with generalized parity restrict the bare QCD
angle to a CP-conserving value and relate the physical strong-CP phase
\(\bar\theta\) to a CP-odd parity-breaking parameter \(\eps\). We show that,
in the Dirac-neutrino realization, imposing a sectorial reality condition on
the Dirac lepton Yukawa matrices turns observable leptonic CP violation from
an independent input into a branch-dependent prediction. Parity reconstructs
the right-handed leptonic mixing matrix, whereas leptonic reality requires it
to be rephasing-equivalent to the complex conjugate of the left-handed one.
For generic three-generation Yukawas, compatibility is equivalent to the
vanishing of a single Jarlskog-type CP-odd invariant.
In the physical Dirac hierarchy, for fixed oscillation
data, mass ordering, and discrete leptonic branch, compatibility determines
the PMNS Jarlskog invariant as a function of the lightest neutrino mass and
\(\eps\). We derive its leading analytic
behavior and obtain the nonlinear compatibility branches numerically.
In compressed mixed-sign Dirac-neutrino spectra, small signed mass sums can
amplify a tiny parity deformation into order-one values of the normalized
PMNS Jarlskog invariant, including maximal CP violation. The corresponding
quark reconstruction independently provides a calculable, branch-dependent
conversion between \(\eps\) and \(\bar\theta\). Together, the leptonic and
quark relations define a family of correlations among leptonic CP violation,
the absolute neutrino mass scale, and strong CP.

\end{abstract}

\maketitle

\section{Introduction}

CP violation appears in several apparently distinct sectors of particle
physics. In the quark sector, it is established through the phase of the
Cabibbo--Kobayashi--Maskawa (CKM) matrix
\cite{Cabibbo:1963yz,Kobayashi:1973fv,ParticleDataGroup:2026aaa}.
Quantum chromodynamics (QCD) admits an additional CP-odd parameter,
the physical strong-CP angle \(\bar\theta\) \cite{tHooft:1976rip}, whose
smallness is constrained by searches for the neutron electric dipole
moment (EDM) \cite{Crewther:1979pi,Abel:2020pzs}.
In the lepton sector, mixing is described by the
Pontecorvo--Maki--Nakagawa--Sakata (PMNS) matrix
\cite{Pontecorvo:1957cp,Maki:1962mu}, but its Dirac phase \(\dCP\)
remains experimentally undetermined
\cite{T2K:2025wet}.
Relating these quantities would turn otherwise independent measurements
into a test of a common organizing principle.

Left--right (LR) symmetry with generalized parity provides a natural setting
for such a connection
\cite{Pati:1974yy,Mohapatra:1974gc,Senjanovic:1975rk,Senjanovic:1978ev}.
When parity is exact, the quark Yukawa matrices are Hermitian and the bare
QCD angle is restricted to a CP-conserving value, allowing the physical
strong-CP phase to vanish in the symmetric limit
\cite{Mohapatra:1978fy,Beg:1978mt,Barr:1991qx}. After spontaneous parity
breaking, the complex bidoublet vacuum introduces a CP-odd parameter
\(\eps\) that controls the departure of the fermion mass matrices from
Hermiticity.

The parity structure reconstructs the right-handed quark mixing matrix
\(V_R^q\) \cite{Zhang:2007da,Senjanovic:2014pva,Senjanovic:2015yea}.
The same reconstruction makes the induced strong-CP phase calculable
\cite{Maiezza:2014ala,Senjanovic:2015yea}.
On the conventional quark branch one recovers the familiar leading estimate
\(\bar\theta\sim (m_t/2m_b)\eps\).
The neutron-EDM limit then restricts \(|\eps|\) to a few
\(\times\,10^{-12}\)
\cite{Maiezza:2014ala,Bertolini:2019out}.
Discrete quark parity signs modify this conversion and therefore the
corresponding bound.

In the Dirac lepton sector, parity analogously determines the right-handed
leptonic mixing matrix in terms of the charged-lepton masses, the
light-neutrino spectrum, the observed PMNS matrix, and \(\eps\), up to
discrete sign assignments \cite{Tello:2026ine}.
Under parity alone, however, the observed left-handed phase \(\dCP\)
remains an input to this reconstruction.

We show that imposing an additional sectorial \(\CPell\) reality condition
on the Dirac lepton Yukawa matrices turns this reconstruction into a
branch-dependent prediction for observable leptonic CP violation. In
mixing-matrix language, the parity-reconstructed \(V_R^\ell\) must then be
rephasing-equivalent to \(V_L^*\). This requirement is not satisfied for
arbitrary low-energy data. For fixed mass ordering, mixing angles, mass
splittings, \(m_{\rm lightest}\), \(\eps\), and discrete leptonic branch,
it selects the compatible values of \(\dCP\).

The small Dirac-neutrino masses make the parity reconstruction especially
sensitive to the neutrino spectrum. The reconstruction exists only within
a finite domain whose characteristic extent in \(|\eps|\) is set
parametrically by \(m_{\rm lightest}/m_\tau\) \cite{Tello:2026ine}.
At the same time, compressed mixed-sign neutrino spectra contain small
signed mass sums that can strongly amplify the leptonic response. Even a
parity deformation small enough to satisfy the strong-CP constraint can
therefore produce order-one values of the normalized PMNS Jarlskog
invariant.

At the Yukawa level, the compatibility problem has a particularly simple
structure. For generic three-generation Yukawas, simultaneous leptonic
reality and parity are equivalent to the vanishing of a single
Jarlskog-type CP-odd invariant. The apparently separate requirements on the
right-handed mixing angles and CP orientation are therefore correlated
consequences of one physical condition. Its Jarlskog projection 
takes a simple form. Writing the parity-reconstructed right-handed
invariant as \(J_R=J_L+\Delta J_P\), while leptonic reality requires
\(J_R=-J_L\), gives
\begin{equation*}
J_L=-\frac12\,\Delta J_P .
\end{equation*}
This identity holds exactly on compatible solutions. Expanding the parity
reconstruction around the CP-conserving solutions gives the leading
analytic response, while the compatibility condition determines the
nonlinear branch structure.

The restriction of the additional reality condition to leptons is
deliberate. Universal \(\Pcal+\CP\) realizations of spontaneous CP
violation have long been studied in the LR framework
\cite{Mohapatra:1977mj,Mohapatra:1978cr,Branco:1982wp,
Chang:1982dp,Ecker:1983hz}.
In the minimal realization, the CKM phase is calculable and linear in
\(\eps\) at leading order \cite{Chang:1982dp}, whereas the strong-CP
constraint requires \(\eps\) to be very small. We therefore retain
generalized parity as the universal LR symmetry while imposing the
additional reality condition only on the Dirac lepton Yukawa matrices.
The quark Yukawas remain general Hermitian matrices, so the observed CKM
phase is not tied to \(\eps\).

The leptonic and quark discrete structures enter at different stages.
Whether the signed neutrino masses add or subtract controls the dominant
enhancement pattern in the leptonic solutions, while the quark parity signs
determine the conversion between \(\eps\) and
\(\bar\theta\). Together, the leptonic compatibility condition and the
quark reconstruction define a family of correlations among leptonic CP
violation, the absolute neutrino mass scale, and strong CP. Measurements of
leptonic CP and the absolute neutrino mass scale can then be combined with
neutron-EDM information to probe the same parity-breaking deformation in
three distinct experimental directions.

\section{Parity reconstruction and the strong-CP map}
\label{sec:setup}

\subsection{Leptonic parity reconstruction}

For the leptonic reconstruction, the spectator \(SU(3)_c\) factor is
suppressed. We work with the left--right gauge structure
\cite{Pati:1974yy,Mohapatra:1974gc}, with generalized parity
\(\Pcal\) \cite{Senjanovic:1975rk,Senjanovic:1978ev},
\begin{equation*}
SU(2)_L\times SU(2)_R\times U(1)_{B-L},
\end{equation*}
in its minimal doublet-breaking realization.
The relevant fermion multiplets are the lepton doublets
\(L_L(2,1,-1)\) and \(L_R(1,2,-1)\). The scalar sector contains the
bidoublet \(\Phi(2,2,0)\) and the LR-breaking doublets
\(\phi_L(2,1,1)\) and \(\phi_R(1,2,1)\). Under parity,
\begin{equation*}
L_L\leftrightarrow L_R,
\qquad
\Phi\leftrightarrow\Phi^\dagger,
\qquad
\phi_L\leftrightarrow\phi_R .
\end{equation*}
The leptonic bidoublet Yukawa interaction is
\begin{equation*}
-\mathcal L_Y^\ell
=
\overline L_L
\left(
Y_1^\ell\Phi+Y_2^\ell\widetilde\Phi
\right)L_R
+\mathrm{H.c.},
\end{equation*}
where \(\widetilde\Phi\equiv\sigma_2\Phi^*\sigma_2\). Parity requires
\(Y_{1,2}^\ell=Y_{1,2}^{\ell\dagger}\) at the scale where parity is imposed.

With this field content, neutrino masses are Dirac at the renormalizable
level \cite{Branco:1978bz}. We assume that higher-dimensional
lepton-number-violating operators
are forbidden or sufficiently suppressed to be irrelevant for the analysis,
so that the light neutrinos may be treated as purely Dirac.

We take
\begin{equation*}
\langle\Phi\rangle
=
v\,\diag(\cos\beta,\sin\beta e^{i\alpha}),
\end{equation*}
and define
\begin{equation}
\eps=\sin\alpha\tan2\beta .
\label{eq:eps_def}
\end{equation}
The resulting Dirac mass matrices are
\begin{align*}
M_\nu
&=
v\left(
\cos\beta\,Y_1^\ell
+
\sin\beta\,e^{-i\alpha}Y_2^\ell
\right),
\\
M_\ell
&=
v\left(
\sin\beta\,e^{i\alpha}Y_1^\ell
+
\cos\beta\,Y_2^\ell
\right).
\end{align*}

Using \(Y_{1,2}^\ell=Y_{1,2}^{\ell\dagger}\), the exact parity constraints
may equivalently be written as
\begin{align}
M_\nu-M_\nu^\dagger
&=
i\eps\left(
e^{i\alpha}\tan\beta\,M_\nu-M_\ell
\right),
\label{eq:full_parity_Mnu}
\\
M_\ell-M_\ell^\dagger
&=
i\eps\left(
M_\nu-e^{-i\alpha}\tan\beta\,M_\ell
\right).
\label{eq:full_parity_Mell}
\end{align}
The charged-lepton and Dirac-neutrino mass matrices are therefore in general
non-Hermitian for nonzero \(\eps\), and the two equations form a coupled
parity-reconstruction problem. Their exact formulation in terms of the
mass-weighted mismatches \(X_\nu\) and \(X_\ell\) is given in
Ref.~\cite{Tello:2026ine}.

We denote the positive mass matrices by
\begin{equation*}
m_\nu=\diag(m_{\nu_1},m_{\nu_2},m_{\nu_3}),
\quad
m_\ell=\diag(m_e,m_\mu,m_\tau),
\end{equation*}
with \(m_{\nu_i}>0\) and \(m_\alpha>0\) \((\alpha=e,\mu,\tau)\).
The discrete parity signs are collected in
\begin{equation*}
S_\nu=\diag(s_{\nu_1},s_{\nu_2},s_{\nu_3}),\qquad S_e=\diag(s_e,s_\mu,s_\tau),
\end{equation*}
with \(s_{\nu_i} =\pm1\) and \(s_\alpha=\pm1\). We define the signed mass matrices
\begin{equation*}
\hat m_\nu=S_\nu m_\nu,
\qquad
 \hat m_\ell=S_e m_\ell.
\end{equation*}
Throughout, unhatted masses are positive, while hatted masses include the
corresponding branch signs. The hatted eigenvalues do not represent negative
physical masses. Their signs label distinct parity-reconstruction branches
and cannot all be removed while keeping the parity basis and the reconstructed
right-handed mixing matrix fixed.

To fix the diagonalization conventions, for each Dirac fermion species
\(f\) we write
\begin{equation}
M_f
=
U_L^{(f)}m_fU_R^{(f)\dagger},
\quad
U_f\equiv U_L^{(f)\dagger}U_R^{(f)},
\quad
X_f\equiv U_fm_f .
\label{eq:Dirac_mismatch_definition}
\end{equation}
Here \(U_f\) measures the mismatch between the left- and right-handed
diagonalizations, while \(X_f\) is its mass-weighted form. Let \(V_L\)
denote the observed PMNS matrix. The left- and right-handed leptonic mixing
matrices are consequently related by
\begin{equation}
V_L
=
U_L^{(\ell)\dagger}U_L^{(\nu)},
\qquad
V_R^\ell
=
U_R^{(\ell)\dagger}U_R^{(\nu)}
=
U_\ell^\dagger V_LU_\nu .
\label{eq:general_leptonic_mismatch}
\end{equation}

For the physical Dirac spectrum \(m_{\nu_i}\ll m_\alpha\), the coupled
reconstruction simplifies \cite{Tello:2026ine}. The neglected neutrino
term is suppressed by \(\mathcal O(\tan\beta\,m_\nu/m_\ell)\), while
off-diagonal charged-lepton corrections are
\(\mathcal O(\eps\,m_\nu/m_\ell)\). The remaining diagonal charged-lepton
corrections amount to external rephasings. For sub-eV neutrino masses,
\(m_\nu/m_e\lesssim10^{-6}\). We therefore use
\begin{equation}
V_R^\ell=S_eV_LU_\nu,
\qquad
X\equiv X_\nu=U_\nu m_\nu .
\label{eq:VRreconstruction}
\end{equation}

Defining \(H\equiv V_L^\dagger\hat m_\ell V_L\), the reduced parity
reconstruction is governed by
\begin{equation}
X^2=m_\nu^2+i\eps HX .
\label{eq:master}
\end{equation}
Physical solutions require \(U_\nu=Xm_\nu^{-1}\) to be unitary.
Unitarity also restricts the solutions of Eq.~\eqref{eq:master} to a
finite domain with characteristic scale
\(|\eps|_{\max}\sim m_{\rm lightest}/m_\tau\).

The analytic and numerical parity reconstruction below uses this physical
hierarchy reduction. Its nonlinear solutions are obtained nonperturbatively
in \(\eps\), while statements made directly at the Yukawa level do not rely
on the reduction.

At \(\eps=0\), the physical solutions reduce to
\begin{equation*}
U_\nu=S_\nu,
\qquad
X=S_\nu m_\nu=\hat m_\nu ,
\end{equation*}
which identifies the signed neutrino branches.

With parity alone, \(V_L\), including \(\dCP\), is an input, while
\(V_R^\ell\) is obtained from the reconstruction. The signed neutrino
branches are important because, in compressed mixed-sign spectra, small
signed mass sums can amplify the response to \(\eps\)
\cite{Tello:2026ine}. This appears through explicit denominators in the
ordinary small-\(\eps\) expansion and through resummed phases in the
quasi-degenerate regime.

\subsection{Quark parity branches and strong CP}
\label{subsec:quark_strong_CP}

The same parity reconstruction applies to the quark sector
\cite{Senjanovic:2014pva,Senjanovic:2015yea}.
We now construct the branch-dependent map between \(\eps\) and the induced
strong-CP phase. For this purpose we require only the corresponding
mass-weighted mismatches,
\begin{equation}
X_u=U_um_u,
\qquad
X_d=U_dm_d .
\end{equation}

We collect the
discrete quark parity signs in
\begin{equation}
S_u=\diag(s_{u_1},s_{u_2},s_{u_3}),
\quad
S_d=\diag(s_{d_1},s_{d_2},s_{d_3}),
\end{equation}
with \(s_{u_i},s_{d_i}=\pm1\), and denote the corresponding quark parity
branch collectively by \(S\equiv(S_u,S_d)\).

We use the convention
\begin{equation*}
\bar\theta
=
\theta_{\rm QCD}
+
\arg\det(M_uM_d)
\qquad(\mathrm{mod}\ 2\pi).
\end{equation*}
Since
\(\arg\det M_f=-\arg\det X_f\), this becomes
\begin{equation}
\bar\theta
=
\theta_{\rm QCD}
-
\arg\det(X_uX_d).
\label{eq:bartheta_X}
\end{equation}

The discrete origin of each branch at \(\eps=0\) must be kept explicit.
 
At the parity-symmetric point, \(\eps=0\),
\begin{equation*}
X_u=S_um_u,
\qquad
X_d=S_dm_d,
\end{equation*}
and therefore
\begin{equation*}
\bar\theta_S(0)
=
\theta_{\rm QCD}
-
\arg\det(S_uS_d)
\qquad(\mathrm{mod}\ 2\pi).
\end{equation*}
The parity-invariant values of the physical strong phase are
\(\bar\theta=0\) and \(\pi\), of which only the former is
phenomenologically viable for the observed quark spectrum
\cite{Vecchi:2025qie}. For each parity-sign class, we therefore choose the
parity-invariant value \(\theta_{\rm QCD}=0\) or \(\pi\) such that
\(\bar\theta_S(0)=0\). We call these the matched branches,
\begin{equation}
\theta_{\rm QCD}
-
\arg\det(S_uS_d)
=
0
\qquad(\mathrm{mod}\ 2\pi).
\label{eq:quark_discrete_matching}
\end{equation}

For nonzero \(\eps\), the parity reconstruction deforms each branch as
\begin{equation}
\bar\theta_S
=
\bar\theta_S(0)
+
C_q^{(S)}\,\eps
+
\OO(\eps^2),
\label{eq:bartheta_branch}
\end{equation}
where, with the convention of Eq.~\eqref{eq:bartheta_X},
\begin{equation}
C_q^{(S)}
=
\frac12\sum_{i,j}
s_{u_i}s_{d_j}
\left|(V_{\rm CKM})_{ij}\right|^2
\left(
\frac{m_{u_i}}{m_{d_j}}
-
\frac{m_{d_j}}{m_{u_i}}
\right).
\label{eq:Cq_branch}
\end{equation}
All quantities entering \(C_q^{(S)}\) are evaluated at a common
renormalization scale, which we denote by \(\mu_P\).
The parity-induced strong-CP response has been studied in
Refs.~\cite{Maiezza:2014ala,Senjanovic:2015yea,Bertolini:2019out}.
On the matched branches,
\begin{equation}
\bar\theta_S
=
C_q^{(S)}\,\eps
+\OO(\eps^2).
\label{eq:bartheta_matched}
\end{equation}
The familiar estimate
\(\bar\theta\sim(m_t/2m_b)\eps\) captures the dominant
third-generation scaling on the conventional branch. The numerical
normalization instead uses the complete coefficient in
Eq.~\eqref{eq:Cq_branch}, including all flavor channels.

The current neutron-EDM limit is
\(|d_n|<1.8\times10^{-26}\,e\,{\rm cm}\) at \(90\%\) C.L.
\cite{Abel:2020pzs}.
Up to hadronic uncertainties
\cite{Pospelov:2005pr,Engel:2013lsa}, this corresponds to the standard
estimate
\begin{equation}
|\bar\theta|\lesssim10^{-10}.
\end{equation}
In converting this estimate into a constraint on \(\eps\), we assume that
no additional mechanism, such as Peccei--Quinn relaxation
\cite{Peccei:1977hh,Peccei:1977ur}, removes or otherwise suppresses the
parity-induced \(\bar\theta\), and that its neutron-EDM contribution is not
cancelled by other CP-odd sources. 

On matched branches for which the linear term dominates, this gives
\begin{equation}
|\eps|
\lesssim
\frac{10^{-10}}{|C_q^{(S)}|}.
\label{eq:epsilon_quark_branch_bound}
\end{equation}
The strong-CP constraint on \(\eps\) is therefore branch dependent.

\section{Sectorial leptonic reality and compatibility}
\label{sec:CPcondition}
 
We impose an additional \(\CPell\), or reality, condition in the Dirac
lepton Yukawa sector, while keeping generalized parity as the LR symmetry.
In the present work, \(\CPell\) denotes the sectorial condition that the
Dirac lepton Yukawa matrices \(Y_{1,2}^\ell\) can be made simultaneously
real in a parity basis. This condition is imposed at the same scale as the
parity Yukawa relations.

Parity requires
\[
Y_{1,2}^\ell=Y_{1,2}^{\ell\dagger},
\]
so compatibility with \(\CPell\) is the question of whether two Hermitian
three-generation Yukawa matrices can be made simultaneously real by a
common flavor-basis transformation preserving the parity form. 

For generic nondegenerate Yukawas, this simultaneous-reality obstruction can
be characterized using the standard three-generation Jarlskog commutator
construction \cite{Jarlskog:1985ht,Jarlskog:1985cw}. For the leptonic
Yukawa pair, the corresponding CP-odd invariant is
\begin{equation}
\mathcal I_Y
\equiv
\frac{1}{6i}
\Tr\!\left(
[Y_1^\ell,Y_2^\ell]^3
\right).
\label{eq:IY_full}
\end{equation}
Indeed, in a basis in which
\(Y_1^\ell=\diag(y_1,y_2,y_3)\),
\begin{equation*}
\mathcal I_Y
=
(y_1-y_2)(y_2-y_3)(y_3-y_1)\,
\Im\!\left[
(Y_2^\ell)_{12}
(Y_2^\ell)_{23}
(Y_2^\ell)_{31}
\right].
\end{equation*}
Two diagonal rephasings remove two of the three off-diagonal phases of
\(Y_2^\ell\), while the phase of their cyclic product is invariant.
Consequently, for generic three-generation Yukawas,
\begin{equation}
\Pcal+\CPell\ \text{compatibility}
\quad\Longleftrightarrow\quad
\mathcal I_Y=0 .
\label{eq:IY_compatibility}
\end{equation}
Thus compatibility reduces to a single condition in the full Dirac Yukawa
system.

More general weak-basis-invariant conditions for CP conservation in the
unrestricted LR fermion mass sector were derived in Ref.~\cite{Branco:1986pr}.
For the parity-restricted Hermitian Yukawa pair considered here,
the simultaneous-reality obstruction reduces to the single
three-generation commutator condition above.

When Eq.~\eqref{eq:IY_compatibility} holds, a parity basis may be chosen in
which \(Y_{1,2}^\ell\) are real symmetric. After the bidoublet acquires its
complex vev, the Dirac charged-lepton and neutrino mass matrices are
therefore complex symmetric. For a symmetric matrix \(M_f^T=M_f\), the
left- and right-handed diagonalizers introduced in
Eq.~\eqref{eq:Dirac_mismatch_definition} may be chosen such that
\begin{equation}
U_R^{(f)}=U_L^{(f)*}K_f,
\end{equation}
where \(K_f\) is diagonal and unitary. The phases in \(K_f\) reflect the
freedom to rephase the Dirac mass eigenstates and carry no independent
mixing information. Applying this relation to charged leptons and neutrinos
gives
\begin{equation}
\VRl=K_e^\dagger \VL^*K_\nu .
\label{eq:CP_relation}
\end{equation}

The same algebraic mixing relation arises whenever the relevant mass
matrices are symmetric, including universal \(\Pcal+\CP\) constructions
\cite{Chang:1982dp,Branco:1982wp,Ecker:1983hz} and when generalized charge
conjugation \(\Ccal\) is chosen instead of parity
\cite{Maiezza:2010ic}. Here, generalized parity remains the LR symmetry,
while \(\CPell\) is an additional restriction imposed only on the Dirac
lepton Yukawa sector.

The invariant formulation makes the parameter counting transparent. The
general parity theory contains ten independent continuous flavor parameters:
six fermion masses, three mixing angles, and one Dirac phase. Imposing
\(\CPell\) restricts the two Hermitian lepton Yukawa matrices to be
simultaneously real and reduces this number to nine. The restricted
\(\Pcal+\CPell\) benchmark therefore imposes one physical relation among
the low-energy flavor observables.
Once the masses, mixing angles, and scalar-sector parameters are specified,
that relation determines the compatible values of \(\dCP\) on each discrete
branch.

In mixing language, the same single condition appears as several
requirements. The two mixing matrices must have identical moduli, while
their Jarlskog invariants, denoted by \(J_L\) and \(J_R\), must have
opposite signs:
\begin{equation}
|V_R^\ell|^2=|V_L|^2,
\qquad
J_R=-J_L .
\end{equation}
Parity, however, has already correlated the right- and left-handed mixing
parameters. The modulus and Jarlskog requirements are therefore different
mixing-space projections of the single Yukawa obstruction
\(\mathcal I_Y\).

For the analytic treatment, denote by \(V_R^P\) the right-handed matrix
reconstructed from parity alone. Its Jarlskog invariant may be written as
\begin{equation}
J_R
=
J_L+\Delta J_P,
\label{eq:JR_deltaJP}
\end{equation}
and compatibility therefore gives
\begin{equation}
J_L=-\frac12\,\Delta J_P .
\label{eq:J_master_condition}
\end{equation}
This Jarlskog identity is also exact in the full Dirac system. We use
the physical hierarchy introduced in Sec.~\ref{sec:setup} to evaluate the
parity response.
 
\section{First-order prediction for the leptonic Jarlskog invariant}
\label{sec:firstorder}

At \(\eps=0\), the parity solution \(U_\nu=S_\nu\) gives
\(V_R^P=S_eV_LS_\nu\). Compatibility with
Eq.~\eqref{eq:CP_relation} then requires \(V_L\sim V_L^*\). For generic
three-generation mixing, the two CP-conserving origins are
\begin{equation}
\dCP=\delta_0,
\qquad
\delta_0=0,\pi .
\label{eq:CP_origins}
\end{equation}
The nonzero-\(\eps\) compatibility branches emerge from these points.
 
For a fixed sign branch
\((S_e,S_\nu)\), the first-order leptonic parity reconstruction of
Ref.~\cite{Tello:2026ine} is
\begin{equation}
V_R^P
=
S_eV_L
\left(
1+i\eps B
\right)
S_\nu
+\OO(\eps^2),
\label{eq:VR_firstorder_general}
\end{equation}
where
\begin{equation}
B_{ij}
=
\frac{(V_L^\dagger\hat m_\ell V_L)_{ij}}
{\hat m_{\nu_i}+\hat m_{\nu_j}} .
\label{eq:B_general}
\end{equation}
The diagonal entries of \(B\) correspond to column rephasings and do not
affect the Jarlskog invariant. Only the off-diagonal entries contribute.

We now expand around \(\delta_0\),
\begin{equation}
\dCP=\delta_0+\eta,
\qquad
\eta=\OO(\eps).
\label{eq:delta_expansion}
\end{equation}
At \(\dCP=\delta_0\) we use the real PMNS representative
\begin{equation*}
V_L(\delta_0)=O,
\qquad
O\in SO(3).
\end{equation*}
Since the parity deformation starts at \(\OO(\eps)\), its dependence on the
\(\OO(\eps)\) displacement \(\eta\) enters only at higher order. The
quantities multiplying the explicit factor of \(\eps\) may therefore be
evaluated at \(V_L=O\). Defining
\begin{equation}
B^{(0)}_{ij}
=
\frac{(O^T\hat m_\ell O)_{ij}}
{\hat m_{\nu_i}+\hat m_{\nu_j}},
\label{eq:B_firstorder}
\end{equation}
the reconstructed matrix becomes
\begin{equation}
V_R^P
=
S_e
\left[
V_L+i\eps\,OB^{(0)}
\right]
S_\nu
+\OO(\eps^2).
\label{eq:VR_firstorder_projected}
\end{equation}
The first term carries the left-handed invariant \(J_L\), while the second
gives the leading parity-induced change \(\Delta J_P\).

Linearizing the Jarlskog invariant
\begin{equation*}
J[V]
=
\Im\!\left(
V_{e1}V_{\mu2}V_{e2}^*V_{\mu1}^*
\right)
\end{equation*}
under the deformation \(i\eps OB^{(0)}\) gives the three response factors
\begin{equation}
\begin{aligned}
n_{12}&=O_{e3}O_{\mu3}O_{\tau3},\\
n_{13}&=O_{e2}O_{\mu2}O_{\tau2},\\
n_{23}&=O_{e1}O_{\mu1}O_{\tau1}.
\end{aligned}
\label{eq:n_def}
\end{equation}
For compactness, we define
\begin{equation*}
\mathbf n
\equiv
(n_{12},n_{13},n_{23}),
\qquad
\mathbf v_P
\equiv
(B^{(0)}_{12},B^{(0)}_{13},B^{(0)}_{23}).
\end{equation*}
The external sign matrices in
Eq.~\eqref{eq:VR_firstorder_projected} do not affect the invariant, and the
parity reconstruction therefore gives
\begin{equation*}
J_R
=
J_L
+
\eps\,
\mathbf n\cdot\mathbf v_P
+\OO(\eps^2).
\end{equation*}

The second term is the leading parity-induced contribution
\(\Delta J_P^{(1)}\). On compatible solutions
\(J_L=J_{\rm CP}^{\ell}\), and Eq.~\eqref{eq:J_master_condition} gives
\begin{equation}
\Jcp
=
-\frac{\eps}{2}\,
\mathbf n\cdot\mathbf v_P
+\OO(\eps^3).
\label{eq:J_firstorder_vector}
\end{equation}
The cubic remainder follows from conjugation symmetry. At fixed discrete
signs, complex conjugation maps the compatible branch locally from
\(\eps\) to \(-\eps\) about either CP-conserving point, with
\(J_{\rm CP}^{\ell}(-\eps)=-J_{\rm CP}^{\ell}(\eps)\). Even powers of
\(\eps\) therefore vanish.

Equivalently,
\begin{equation}
\Jcp
=
-\frac{\eps}{2}
\sum_{i<j}
\frac{(O^T\hat m_\ell O)_{ij}}
{\hat m_{\nu_i}+\hat m_{\nu_j}}
\,O_{ek}O_{\mu k}O_{\tau k}
+\OO(\eps^3),
\label{eq:J_firstorder}
\end{equation}
where \(\{i,j,k\}=\{1,2,3\}\).
All mixing-angle dependence is contained in the real matrix \(O\).
An explicit decomposition into charged-lepton response coefficients,
together with the enhanced branch limits, is given in
Appendix~\ref{app:explicit}.

In terms of the standard PMNS phase,
\(\Jcp=\Jmax\sin\dCP\), with
\(\Jmax\equiv c_{12}s_{12}c_{23}s_{23}c_{13}^{2}s_{13}\).
Together with Eq.~\eqref{eq:delta_expansion} and defining
\(\zeta\equiv\cos\delta_0=\pm1\), this gives
\begin{equation}
\dCP
=
\delta_0
-
\frac{\eps}{2\zeta\Jmax}\,
\mathbf n\cdot\mathbf v_P
+\OO(\eps^3).
\label{eq:delta_firstorder}
\end{equation}

Equations~\eqref{eq:J_firstorder_vector}--\eqref{eq:delta_firstorder}
give equivalent forms of the leading compatibility prediction.
Equation~\eqref{eq:J_firstorder_vector} separates the three parity-response
channels, while Eq.~\eqref{eq:J_firstorder} makes the signed-neutrino
denominators explicit. Equation~\eqref{eq:delta_firstorder} gives the 
corresponding PMNS-phase displacement. For fixed masses, mixing
angles, parity breaking, and discrete signs, leptonic CP is therefore no
longer an independent input.

\subsection{Mixed-sign enhancement}
\label{subsec:mixedsign}

The enhancement in Eq.~\eqref{eq:J_firstorder} is controlled by the signed
neutrino denominators. On mixed-sign branches, these sums become mass
differences and can be small in compressed spectra. We illustrate the
\(12\)-enhanced case, which gives the strongest enhancement in the ordinary
expansion.

For \(s_{\nu_1}=+1\) and \(s_{\nu_2}=-1\), with \(s_{\nu_3}\) arbitrary,
retaining the \(12\) channel in Eq.~\eqref{eq:delta_firstorder} and the
dominant \(\tau\)-mass contribution gives the parametric phase shift
\begin{equation}
\left.\dCP\right|_{12}
\sim
\delta_0+
\eps\,
\frac{\hat m_\tau}
{m_{\nu_1}-m_{\nu_2}},
\label{eq:delta_12_scaling}
\end{equation}
up to an order-one angular factor. The explicit angular coefficient and
subleading charged-lepton contributions are given in
Appendix~\ref{app:explicit_firstorder}.

Since
\begin{equation}
\frac{1}{m_{\nu_1}-m_{\nu_2}}
=
-\frac{m_{\nu_1}+m_{\nu_2}}{\Delta m_{21}^2},
\label{eq:solar_difference_identity}
\end{equation}
the response is enhanced as the first two neutrino masses become
increasingly compressed.

With parity alone,
a small signed neutrino denominator enhances the parity-induced
right-handed response. Once leptonic reality is imposed, the left-handed
phase is no longer free: compatibility requires
\(J_L=-\Delta J_P/2\). The spectral enhancement therefore appears as
a large compatible value of \(J_{\rm CP}^{\ell}\) on compressed mixed-sign
branches.

The ordinary expansion assumes
\begin{equation*}
|\eps|\max_{i,j}|B_{ij}|\ll1 .
\end{equation*}
When a mixed-sign denominator becomes sufficiently small, this expansion
ceases to be reliable. The corresponding compressed regime is described by
the resummed quasi-degenerate expansion.

\section{Quasi-degenerate resummation}
\label{sec:QD}

We derive the corresponding response in the complementary
quasi-degenerate (QD) expansion, using the resummed parity reconstruction
of Ref.~\cite{Tello:2026ine}.

We expand around a common neutrino mass scale
\(m_0\simeq m_{\rm lightest}\),
\begin{equation*}
m_{\nu_i}^2=m_0^2(1+\Delta_i),
\qquad
|\Delta_i|\ll1,
\end{equation*}
and define
\begin{equation*}
\Delta=\diag(\Delta_1,\Delta_2,\Delta_3).
\end{equation*}

Since \(J\) is invariant under external row and column rephasings, we remove
the leading charged-lepton signs and phases from the parity solution by
defining
\begin{equation*}
\overline V_R
\equiv
P_\phi^\dagger S_eV_R^P,
\qquad
P_\phi
=
\diag(e^{i\phi_e},e^{i\phi_\mu},e^{i\phi_\tau}).
\end{equation*}
The quasi-degenerate parity solution then takes the form
\begin{equation}
\overline V_R
=
V_L
+
KV_L
-\frac12\,V_L\Delta
+\OO(\Delta^2),
\label{eq:VR_QD_general}
\end{equation}
where
\begin{equation}
K_{\alpha\beta}
=
\frac{
(V_L\Delta V_L^\dagger)_{\alpha\beta}
}{
1+e^{i(\phi_\alpha+\phi_\beta)}
}.
\label{eq:K_QD_general}
\end{equation}

The resummed phases are
\begin{equation}
e^{i\phi_\alpha}
=
\sigma_\alpha
\sqrt{
1-\left(
\frac{\eps\,\hat m_\alpha}{2m_0}
\right)^2
}
+
i\,\frac{\eps\,\hat m_\alpha}{2m_0},
\qquad
\alpha=e,\mu,\tau,
\label{eq:QD_phases}
\end{equation}
where \(\hat m_\alpha=s_\alpha m_\alpha\), while
\(\sigma_\alpha=\pm1\) labels the quasi-degenerate phase branch.
The phase solution requires
\(|\eps|m_\alpha\leq2m_0\), with the strongest condition set by \(m_\tau\).

We expand around the same CP-conserving points as in
Sec.~\ref{sec:firstorder},
\begin{equation}
\dCP=\delta_0+\eta,
\qquad
V_L(\delta_0)=O,
\qquad
O\in SO(3).
\end{equation}

For fixed nonzero \(\eps\), a perturbative quasi-degenerate branch has
\(\eta=\OO(\Delta)\).
Since the numerator of \(K[V_L]\) and the term \(V_L\Delta\) are already
first order in the splittings, they may be evaluated at \(O\) to the
required accuracy. Defining
\begin{equation}
A\equiv O\Delta O^T,
\qquad
K^{(0)}_{\alpha\beta}
\equiv
K[O]_{\alpha\beta}
=
\frac{A_{\alpha\beta}}
{1+e^{i(\phi_\alpha+\phi_\beta)}},
\label{eq:K_QD_seed}
\end{equation}
the reconstruction becomes
\begin{equation}
\overline V_R
=
V_L
+
K^{(0)}O
-\frac12\,O\Delta
+\OO(\Delta^2).
\label{eq:VR_QD_projected}
\end{equation}

Its CP-odd part is
\begin{equation}
\Im K^{(0)}_{\alpha\beta}
=
A_{\alpha\beta}\kappa_{\alpha\beta},
\qquad
\kappa_{\alpha\beta}
\equiv
-\frac12
\tan\!\left(
\frac{\phi_\alpha+\phi_\beta}{2}
\right).
\label{eq:kappa_def}
\end{equation}

Linearizing the Jarlskog invariant in the imaginary off-diagonal components
of \(K^{(0)}\) gives the response factors
\begin{equation}
\begin{aligned}
r_{e\mu}&=O_{\tau1}O_{\tau2}O_{\tau3},\\
r_{e\tau}&=O_{\mu1}O_{\mu2}O_{\mu3},\\
r_{\mu\tau}&=O_{e1}O_{e2}O_{e3}.
\end{aligned}
\label{eq:flavour_response}
\end{equation}
We collect the response factors and the corresponding CP-odd deformation
into
\begin{equation*}
\mathbf r
\equiv
(r_{e\mu},r_{e\tau},r_{\mu\tau}),
\qquad
\mathbf v_{\rm QD}
\equiv
(\Im K^{(0)}_{e\mu},\Im K^{(0)}_{e\tau},\Im K^{(0)}_{\mu\tau}).
\end{equation*}

The parity-induced change in the right-handed invariant is therefore
\begin{equation*}
\Delta J_P^{\rm QD}
=
\mathbf r\cdot\mathbf v_{\rm QD}
+\OO(\Delta^2).
\end{equation*}
On compatible solutions, \(J_R=-J_L\), and therefore
\begin{equation}
J_{\rm CP}^{\ell,{\rm QD}}
=
-\frac12\,
\mathbf r\cdot\mathbf v_{\rm QD}
+\OO(\Delta^2).
\label{eq:J_QD_vector}
\end{equation}
In components,
\begin{equation}
J_{\rm CP}^{\ell,{\rm QD}}
=
-\frac12
\sum_{\alpha<\beta}
r_{\alpha\beta}A_{\alpha\beta}
\kappa_{\alpha\beta}
+\OO(\Delta^2).
\label{eq:J_QD_general}
\end{equation}
This is the quasi-degenerate counterpart of
Eq.~\eqref{eq:J_firstorder}. In this regime, the signed denominators of the
ordinary expansion are resummed into the phases \(\phi_\alpha\). The
enhancement occurs when the magnitude of the coefficient
\(\kappa_{\alpha\beta}\) in Eq.~\eqref{eq:kappa_def} becomes large, namely
when
\begin{equation}
1+e^{i(\phi_\alpha+\phi_\beta)}
\simeq0,
\end{equation}
or equivalently when
\(\phi_\alpha+\phi_\beta\simeq\pi\) modulo \(2\pi\).

The result applies while the QD correction remains perturbative and
\(\dCP\) remains close to its CP-conserving value \(\delta_0\).
Outside this regime, the full nonlinear compatibility solution is required.

\subsection{Mixed-sign enhancement in the QD regime}
\label{subsec:QD_mixedsigma}

As a representative mixed QD branch, consider
\((\sigma_e,\sigma_\mu,\sigma_\tau)=(+,+,-)\) and the asymptotic regime
\(|\eps\hat m_\alpha/(2m_0)|\ll1\).
On this branch, both the \(e\tau\) and \(\mu\tau\) phase sums approach
\(\pi\), while the \(e\mu\) pair remains regular. The two enhanced
contributions interfere, and to leading order in the charged-lepton
hierarchy their sum reduces to
\begin{equation}
J_{\rm CP}^{\ell,{\rm QD}}
\simeq
-r_{e\mu}A_{e\mu}
\frac{m_0}{\eps\hat m_\tau}.
\label{eq:J_QD_ppm_leading}
\end{equation}
A derivation, including the interference between the enhanced
\(e\tau\) and \(\mu\tau\) phase pairs and its relation to the ordinary
first-order response coefficients, is given in
Appendix~\ref{app:QDexplicit}.
 
For the measured mixing angles, the atmospheric contribution to
\(A_{e\mu}\) dominates. Writing it in terms of the signed splitting
\(\Delta m_{31}^2=m_{\nu_3}^2-m_{\nu_1}^2\), the corresponding
phase response is

\begin{equation}
\left.\dCP\right|_{\rm QD}^{(++-)}
\sim
\delta_0
+
\frac{\Delta m_{31}^2}
{\eps\,\hat m_\tau m_0},
\label{eq:delta_QD_ppm_scaling}
\end{equation}
up to an order-one angular factor.
Thus the quasi-degenerate phase response scales inversely with
\(\eps m_0\), in contrast to the ordinary mixed-sign enhancement, whose
phase shift grows proportionally to \(\eps m_0\).
 
The inverse dependence in Eq.~\eqref{eq:delta_QD_ppm_scaling} is an
asymptotic behavior of the quasi-degenerate expansion, not a physical
singularity at \(\eps=0\). Toward sufficiently small \(|\eps|\), the
enhanced QD expansion ceases to be valid and the nonlinear
compatibility solution must be used instead.
Conversely, at exact degeneracy, \(\Delta=0\), one has \(A=0\) and the QD
deformation vanishes. In this limit the neutrino mixing parameters
themselves are unphysical. The enhancement therefore occurs away from both
exact degeneracy and vanishing parity breaking. 
  
\section{Numerical results and phenomenology}
\label{sec:phenomenology}

We now determine the nonlinear compatibility branches numerically. For each
mass ordering and leptonic branch, the solutions define a relation among
\(J_{\rm CP}^{\ell}/J_{\max}\), \(m_{\rm lightest}\), and \(|\eps|\).
At fixed \(m_{\rm lightest}\), this relation describes how leptonic CP
changes as \(|\eps|\) varies.
A measured value of \(J_{\rm CP}^{\ell}\) selects regions in the
\((m_{\rm lightest},|\eps|)\) plane. The quark reconstruction in
Eq.~\eqref{eq:bartheta_matched} then provides the branch-dependent
conversion between \(|\eps|\) and \(|\bar\theta|\).
 
\subsection{Numerical setup}
\label{subsec:numerical_inputs}

Within the hierarchy reconstruction used in the numerical analysis, the
condition \(\mathcal I_Y=0\) is equivalent to
\begin{equation}
\Tr\!\left([X^\dagger,H]^3\right)=0,
\qquad
H=V_L^\dagger\hat m_\ell V_L ,
\label{eq:IY_X_zero}
\end{equation}
which imposes one real condition.

We use the NuFIT~v6.1 best-fit oscillation parameters
\cite{Esteban:2024eli,NuFIT:v61} and consider
\begin{equation*}
m_{\rm lightest}=0.01\,{\rm eV},
\qquad
m_{\rm lightest}=0.3\,{\rm eV}.
\end{equation*}
The first represents a light spectrum, while the second is a diagnostic
benchmark chosen to expose the compressed and quasi-degenerate nonlinear
structure. It corresponds approximately to \(m_\beta\simeq0.3\,{\rm eV}\),
below the present direct kinematic bound \cite{KATRIN:2024cdt}, although it
lies outside the baseline \(\Lambda\)CDM region for stable neutrinos
\cite{Elbers:2025vlz}. 

We take \(\dCP\in[0,2\pi)\) and display the invariant ratio
\(J_{\rm CP}^{\ell}/J_{\max}=\sin\dCP\) as a function of \(|\eps|\).
The sign of \(\eps\) is redundant once all charged-lepton sign branches
are included, as detailed in Appendix~\ref{app:numerical_details}.

For the strong-CP shading in the figures, the representative choice
\(\mu_P=10\,{\rm TeV}\), together with the nominal criterion
\(|\bar\theta|<10^{-10}\), gives
\begin{equation*}
|\eps|_{\rm all}\simeq2\times10^{-12},
\qquad
|\eps|_{\rm none}\simeq7.8\times10^{-12}.
\end{equation*}
Below the first value, all matched quark branches satisfy the nominal
strong-CP bound. Between the two values, only a subset satisfies it. 
Above the second value, none do.

The displayed solutions are tree-level benchmarks evaluated at central
low-energy leptonic inputs, not statistical confidence regions. Leptonic RG
evolution to the parity scale is not included and can be relevant for
quasi-degenerate Dirac neutrinos \cite{Lindner:2005as}.
The explicit inputs, compatibility reduction, branch conventions, and quark
normalization are collected in Appendix~\ref{app:numerical_details}.

\subsection{Fixed-mass branch structure}
\label{subsec:fixed_mass_response}

\begin{figure}[t]
\centering
\makebox[\linewidth][c]{%
  \hspace*{2pt}%
  \includegraphics[width=1.02\linewidth]
    {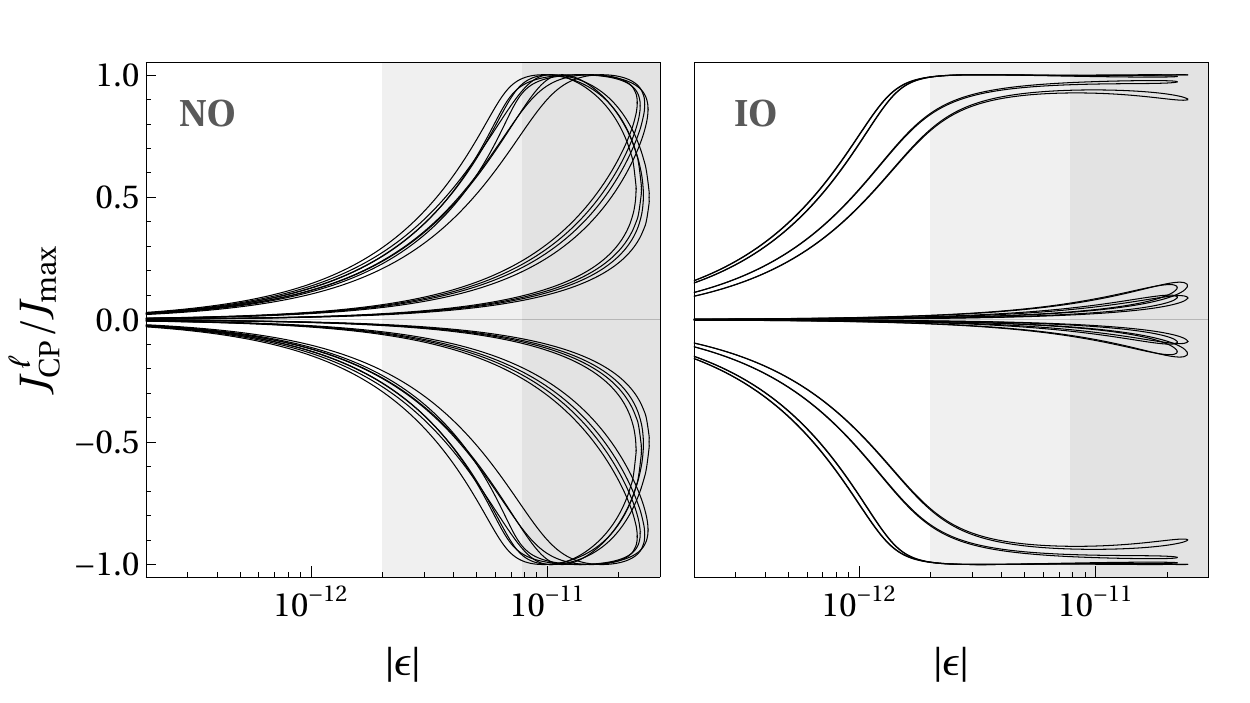}%
  \hspace*{4pt}%
}
\caption{
Numerical fixed-mass solutions at
\(m_{\rm lightest}=0.01\,{\rm eV}\), with NO and IO shown in the left and
right panels, respectively. We show
\(J_{\rm CP}^{\ell}/J_{\max}\) as a function of \(|\eps|\).
All leptonic discrete branches are included, and the complete set is
symmetric under \(J_{\rm CP}^{\ell}\to-J_{\rm CP}^{\ell}\).
In NO, the solar, atmospheric, and same-sign families remain visibly
separated, whereas in IO the atmospheric and same-sign branches are more
closely grouped. 
The vertical shading indicates the quark-branch dependence of
the neutron-EDM constraint: 
all matched quark branches satisfy the nominal strong-CP bound below the
first boundary, only a subset does so in the intermediate region, and none
satisfy it beyond the second.
}
\label{fig:J_response_light}
\end{figure}

Figure~\ref{fig:J_response_light} shows the nonlinear solution set at
\(m_{\rm lightest}=0.01\,{\rm eV}\). In this light-mass regime, the evolution
is strongly branch dependent. In NO, branches dominated by the
solar-enhanced \(12\) channel grow first, those
involving the atmospheric mass splitting become sizable at larger
\(|\eps|\), and the same-sign branches remain weak until
close to the physical endpoint. 

In IO, the \(12\)-enhanced branches rise more rapidly and approach
saturation at smaller \(|\eps|\). By contrast, branches involving the
atmospheric mass splitting remain much weaker and cluster close to the
same-sign family. This reflects the stronger \(12\) compression in IO,
while the atmospheric denominators remain of comparable size in the two
orderings and their contributions partially cancel.

Within the strong-CP-compatible region, sizable CP violation is
dominated by the \(12\)-enhanced branches. The atmospheric
families remain modest, particularly in IO.

\begin{figure}[t]
\centering
\makebox[\linewidth][c]{%
  \hspace*{2pt}%
  \includegraphics[width=1.02\linewidth]
    {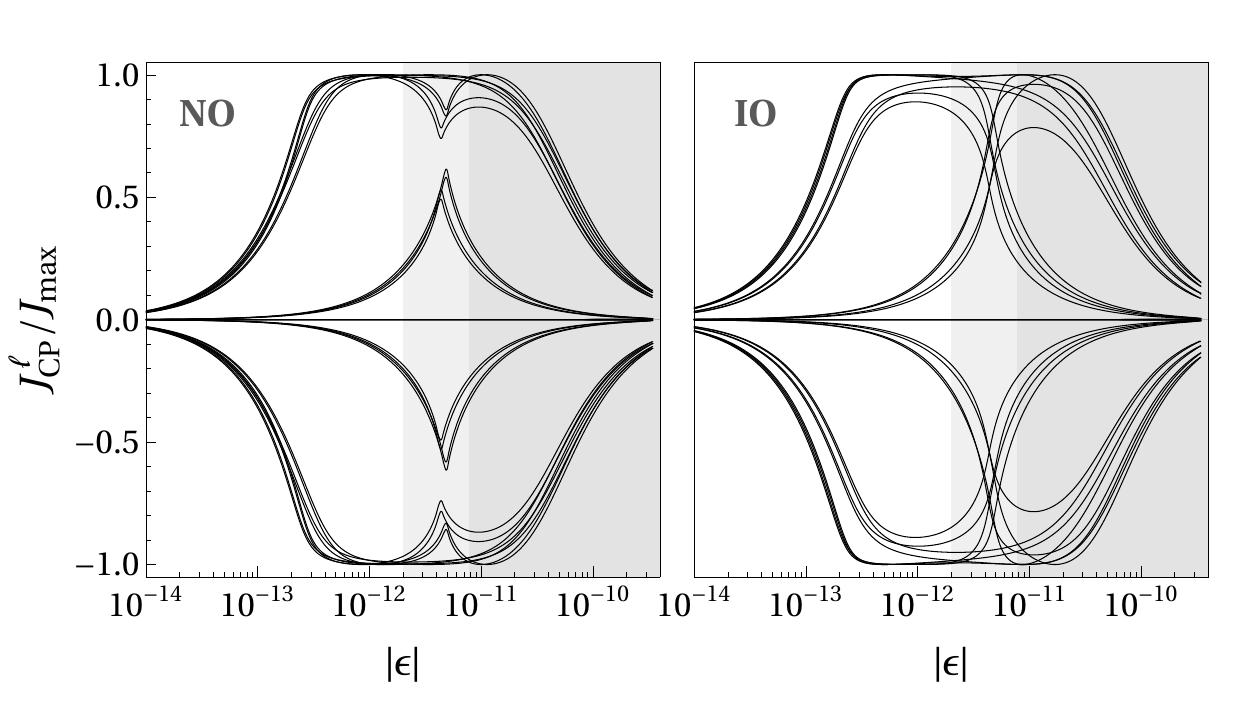}%
  \hspace*{4pt}%
}
\caption{
Numerical fixed-mass solutions at the diagnostic benchmark
\(m_{\rm lightest}=0.3\,{\rm eV}\), which lies outside the baseline
\(\Lambda\)CDM region for stable neutrinos. NO and IO are shown in the left
and right panels, respectively. We show
\(J_{\rm CP}^{\ell}/J_{\max}\) as a function of \(|\eps|\). 
The compressed spectrum shifts the mixed-sign enhancement to substantially
smaller \(|\eps|\), allowing sizable values, including maximal leptonic CP
violation, within the strong-CP-compatible region, while the same-sign
family remains weak. The vertical shading follows the convention of
Fig.~\ref{fig:J_response_light}.
}
\label{fig:J_response_QD}
\end{figure}

Figure~\ref{fig:J_response_QD} shows the nonlinear solutions at
\(m_{\rm lightest}=0.3\,{\rm eV}\). As the spectrum becomes compressed,
the mixed-sign enhancement sets in at smaller \(|\eps|\) relative to the
parity-reconstruction endpoint. On the ordinary rising flanks, the
\(12\)-enhanced branches can reach sizable values, including maximal CP
violation, within the strong-CP-compatible region. The atmospheric families
also become important at smaller \(|\eps|\) than in the light-mass
benchmark, while the same-sign family remains weak. As expected in the
compressed regime, the distinction between NO and IO becomes less
pronounced.

These fixed-mass plots are multibranch solution sets, not unique
phenomenological predictions. At fixed
\(J_{\rm CP}^{\ell}/J_{\max}\), the individual branches generically give
discrete solutions for \(|\eps|\). How sharply these solutions can be
localized once finite experimental phase intervals are included is
illustrated below by the compatibility bands.

Appendix~\ref{app:delta_branches} shows the same solutions directly
in terms of \(\dCP\). This representation resolves phase evolution that is
compressed by the invariant form near maximal CP violation.

\subsection{Comparison with the analytic limits}
\label{subsec:analytic_matching}

The nonlinear solutions give the complete branch evolution. We now compare
them with the analytic limits derived above, using representative NO branches
for clarity. The aim is to show how the analytic expressions describe
different parts of the numerical curves, rather than to compare the two mass
orderings.

\begin{figure}[t]
\centering
\makebox[\linewidth][c]{%
  \hspace*{2pt}%
  \includegraphics[width=1.02\linewidth]{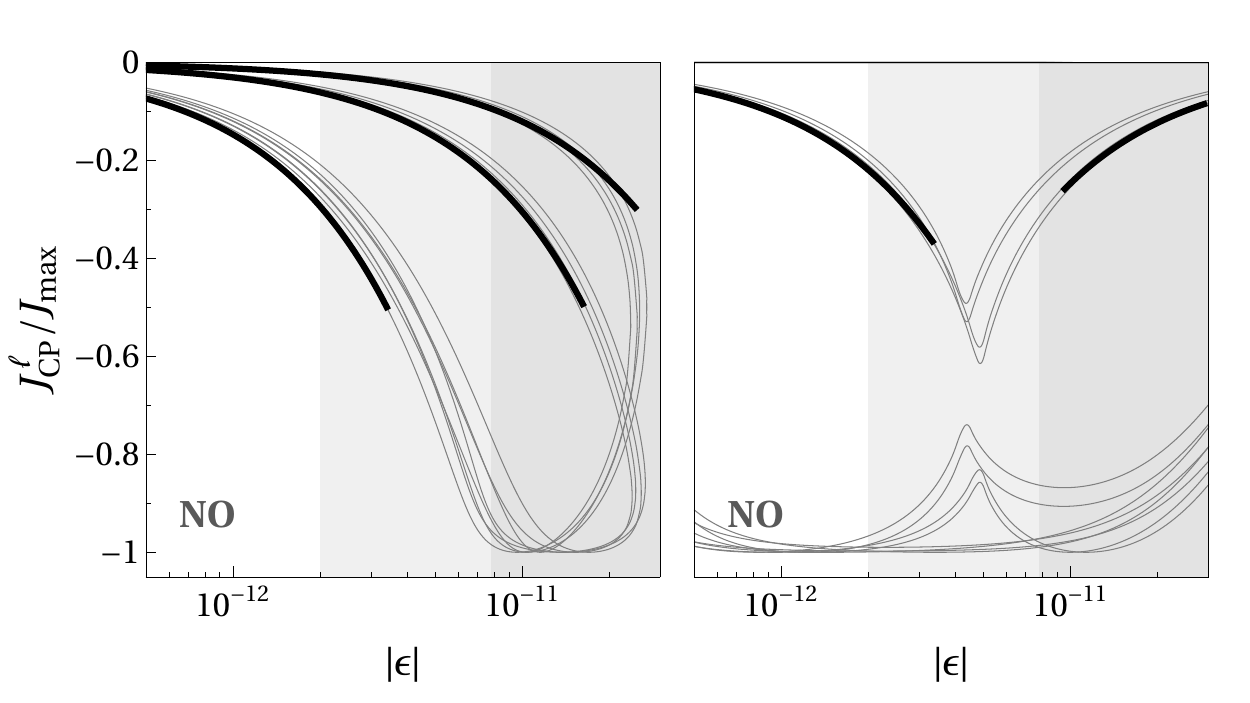}%
  \hspace*{4pt}%
}
\caption{
Comparison of the numerical and analytic descriptions for representative
NO compatibility branches. Only the negative-\(J_{\rm CP}^{\ell}\) half of
the symmetric solution set is shown. Thin gray lines denote the nonlinear
numerical solutions and black curves the corresponding analytic
predictions. The left panel, at \(m_{\rm lightest}=0.01\,{\rm eV}\),
compares the first-order description of same-sign, atmospheric, and
solar-enhanced branches. The right panel follows an atmospheric branch at
\(m_{\rm lightest}=0.3\,{\rm eV}\) through the portion described by the
first-order expansion and the returning portion described by the QD
expansion. Analytic curves are shown only for
\(\rho_{\rm FO,QD}\leq0.5\) and
\(|\eta_{\rm FO,QD}|\leq0.5\).
The vertical shading follows the strong-CP convention of
Fig.~\ref{fig:J_response_light}.
}
\label{fig:analytic_matching}
\end{figure}

Figure~\ref{fig:analytic_matching} overlays the nonlinear numerical
solutions with the first-order and quasi-degenerate predictions. We show
each analytic curve only where both the matrix correction and the
displacement from the CP-conserving origin remain small. We use
\begin{equation*}
\rho_{\rm FO}
\equiv
|\eps|\max_{i,j}|B^{(0)}_{ij}|,
\qquad
\rho_{\rm QD}
\equiv
\max_{\alpha,\beta}|K^{(0)}_{\alpha\beta}|,
\end{equation*}
and require
\begin{equation*}
\rho_{\rm FO,QD}\leq0.5,
\qquad
|\eta_{\rm FO,QD}|\leq0.5.
\end{equation*}
These are practical limits used to display the truncated formulas. They do
not restrict the existence of the nonlinear branches. The first-order and
QD curves are obtained from Eqs.~\eqref{eq:J_firstorder} and
\eqref{eq:J_QD_general}, respectively.

The left panel, at \(m_{\rm lightest}=0.01\,{\rm eV}\), compares the
first-order description of representative same-sign, atmospheric, and
solar-enhanced branches. The right panel, at
\(m_{\rm lightest}=0.3\,{\rm eV}\), follows an atmospheric branch through
the portion described by the first-order expansion and the returning
portion described by the QD expansion.

For the analytic examples in Fig.~\ref{fig:analytic_matching}, we take
\(S_e=(+++)\) and \(\delta_0=\pi\) (\(\zeta=-1\)). The left-panel
same-sign, atmospheric, and solar-enhanced branches use
\(S_\nu=(+++),(++-),(-+-)\), respectively. The atmospheric branch in the
right panel uses \(S_\nu=(++-)\), whose QD continuation has
\((\sigma_e,\sigma_\mu,\sigma_\tau)=(+,+,-)\).

The first-order and QD curves obey, respectively,
\begin{equation}
\frac{J_{\rm CP}^{\ell}}{J_{\max}}
\simeq
-\frac{\eps m_\tau m_0}{\Delta m_{31}^2}\,C_{12}^{\tau},
\qquad
\frac{J_{\rm CP}^{\ell}}{J_{\max}}
\simeq
-\frac{\Delta m_{31}^2}{\eps m_\tau m_0}\,C_{12}^{\tau}.
\label{eq:atm_two_asymptotics}
\end{equation}
For the NO mixing angles used here and \(\zeta=-1\),
\(C_{12}^{\tau}\simeq0.52\).
The reduction of both asymptotic responses to the common coefficient
\(C_{12}^{\tau}\) is detailed in Appendix~\ref{app:explicit}.
The magnitude of the first-order response grows with \(|\eps|\), while that
of the QD response decreases as \(1/|\eps|\). The two expansions therefore
describe opposite sides of the same nonlinear branch. The numerical
solution connects them through the intermediate region, where neither
truncated expansion need be quantitatively accurate.

\subsection{Compatibility bands and mass-scale interpretation}
\label{subsec:compatibility_bands}

To illustrate how a phase measurement constrains the compatibility surface,
we select representative intervals around the current ordering-dependent
preferred regions \cite{Esteban:2024eli,NuFIT:v61},
\begin{align}
\dCP &\in [200^\circ,225^\circ],
\qquad {\rm NO},
\\
\dCP &\in [250^\circ,290^\circ],
\qquad {\rm IO}.
\end{align}
These are illustrative benchmark intervals, not confidence regions. The NO
window probes sloped portions of the solution curves around its present
best-fit region, whereas the IO window covers the neighbourhood of maximal
negative CP violation. Since
\(J_{\rm CP}^{\ell}/J_{\max}=\sin\dCP\) is close to saturation in the
latter case, the wider IO phase interval still corresponds to a 
comparatively narrow range of \(J_{\rm CP}^{\ell}/J_{\max}\).

\begin{figure}[t]
\centering
\makebox[\linewidth][c]{%
  \hspace*{2pt}%
  \includegraphics[width=1.02\linewidth]
    {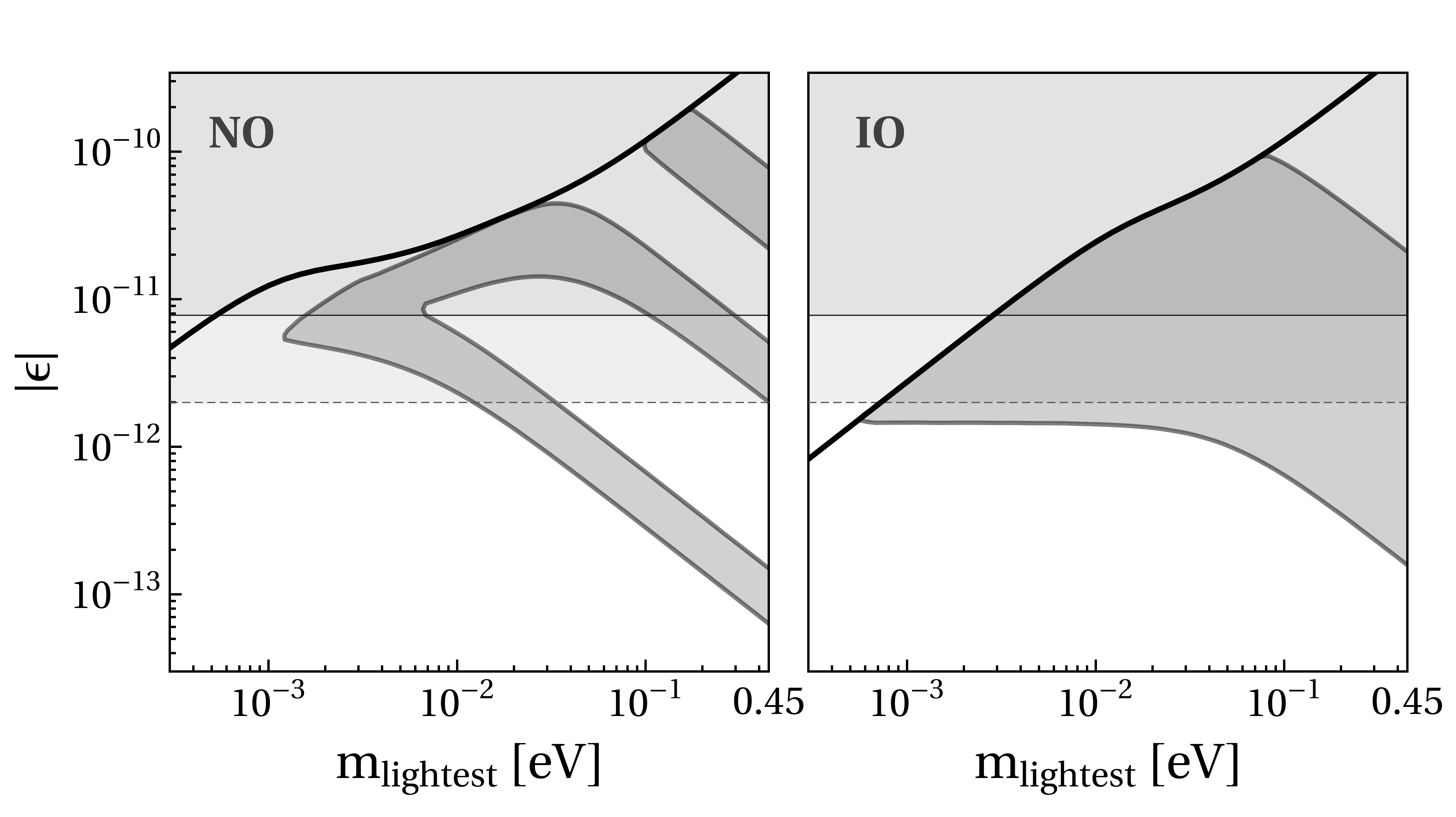}%
  \hspace*{4pt}%
}
\caption{Benchmark compatibility regions in the
\((m_{\rm lightest},|\eps|)\) plane, with NO and IO shown in the left and
right panels, respectively, for
\(\dCP\in[200^\circ,225^\circ]\) in NO and
\(\dCP\in[250^\circ,290^\circ]\) in IO.
The projected regions are the union of all leptonic compatibility solutions
satisfying the respective phase cut.
The black curve marks the upper envelope of the compatibility domain with
\(\dCP\) unrestricted.
The horizontal dashed and solid lines denote
\(|\eps|=2\times10^{-12}\) and \(7.8\times10^{-12}\), respectively.
Below the dashed line all matched quark parity branches satisfy the nominal
neutron-EDM constraint. 
Between the two boundaries only a subset satisfies the nominal constraint,
while above the solid line none do.
}
\label{fig:global_bands}
\end{figure}

For each \(m_{\rm lightest}\), we retain the points on the leptonic branches
whose phase lies inside the selected interval. Scanning over
\(m_{\rm lightest}\) and projecting these solutions onto the
\((m_{\rm lightest},|\eps|)\) plane gives Fig.~\ref{fig:global_bands}.
The quark relation
\(\bar\theta_S\simeq C_q^{(S)}\eps\) then gives the corresponding
strong-CP interpretation for each quark branch. The same phase intervals
are indicated in the direct-\(\dCP\) branch plots of
Appendix~\ref{app:delta_branches}.
 
The selected phase intervals admit compatible solutions down to
\(m_{\rm lightest}\simeq1.2\times10^{-3}\,{\rm eV}\) in NO and
\(6\times10^{-4}\,{\rm eV}\) in IO. In both orderings, substantial parts
of the projected regions lie below
\(|\eps|=7.8\times10^{-12}\) and are therefore compatible with at least one
matched quark branch. Portions of these regions extend below
\(|\eps|=2\times10^{-12}\), where all matched branches satisfy the nominal
strong-CP bound. Above \(7.8\times10^{-12}\), no matched quark branch 
satisfies the nominal bound.

Within the linear matching
\(\bar\theta_S\simeq C_q^{(S)}\eps\), a more stringent neutron-EDM limit
would lower the corresponding \(|\eps|\) thresholds. Compatibility with
the selected phase intervals would then be pushed toward larger
\(m_{\rm lightest}\), and hence toward increasingly compressed neutrino
spectra.

The different shapes mainly reflect where the chosen phase intervals
intersect the multibranch surface. The NO interval mostly selects sloped
regions and therefore gives narrow projected bands. The IO interval lies
near saturation, where the nearly flat response maps a finite phase
interval onto a broader range of \(|\eps|\). These features should not be
interpreted as generic differences between the two mass orderings.

\subsection{Experimental interpretation}
\label{subsec:experimental_interpretation}

The compatibility relation can be tested through three complementary
laboratory observables: leptonic CP violation, the absolute neutrino mass
scale, and the neutron EDM. Oscillation measurements probe the mass ordering
and constrain \(J_{\rm CP}^{\ell}\). Direct kinematic measurements constrain
\(m_{\rm lightest}\) once the ordering and oscillation splittings are
specified. Neutron-EDM searches constrain the parity-breaking parameter
through the quark-sector relation
\(\bar\theta_S\simeq C_q^{(S)}\eps\).
For a fixed mass ordering and discrete branches,
\(J_{\rm CP}^{\ell}\), \(m_{\rm lightest}\), and \(|\bar\theta|\) are
therefore correlated rather than independent.

The next generation of long-baseline experiments, in particular
Hyper-Kamiokande and DUNE, will measure the leptonic phase more precisely
\cite{Hyper-Kamiokande:2025fci,DUNE:2020jqi}, while ESS\(\nu\)SB targets
a high-precision determination of \(\dCP\) \cite{ESSnuSB:2023ogw}.
 
Direct kinematic measurements probe
\begin{equation*}
m_\beta^2
=
\sum_i |(V_L)_{ei}|^2 m_{\nu_i}^2 .
\end{equation*}
KATRIN currently gives
\(m_\beta<0.45\,{\rm eV}\) at \(90\%\) C.L.
\cite{KATRIN:2024cdt}, while Project~8 ultimately targets a sensitivity of
about \(40\,{\rm meV}\) \cite{Project8:2022wqh}. In the compressed regime,
\(m_\beta\simeq m_{\rm lightest}\).

Neutron-EDM searches probe the third direction. The present limit gives the
branch-dependent constraint of
Eq.~\eqref{eq:epsilon_quark_branch_bound}, while n2EDM is designed to reach
sensitivity near \(10^{-27}\,e\,{\rm cm}\)
\cite{n2EDM:2021yah}. Improved EDM sensitivity therefore progressively
restricts the allowed \(|\eps|\) range, with the conversion depending on
the quark parity branch.

We use \(\mathcal B_\ell\) to specify the leptonic parity-sign class and its
CP-conserving origin, and write the compatibility relation as
\[
\frac{J_{\rm CP}^{\ell}}{J_{\max}}
=
F_\ell\!\left(
m_{\rm lightest},|\eps|;
\mathcal B_\ell,\text{ordering}
\right).
\]
Combining this relation with the quark-sector map gives 
\begin{equation}
\frac{J_{\rm CP}^{\ell}}{J_{\max}}
\simeq
F_\ell\!\left(
m_{\rm lightest},
\frac{|\bar\theta|}{|C_q^{(S)}|};
\mathcal B_\ell,\text{ordering}
\right).
\label{eq:final_J_theta_correlation}
\end{equation}
The prediction is therefore a family of branch-labelled correlations in
the three-dimensional space
\((J_{\rm CP}^{\ell},m_{\rm lightest},|\bar\theta|)\).

Measurements of \(J_{\rm CP}^{\ell}\) and the absolute neutrino mass scale
select compatible values of \(|\eps|\), and hence a branch-dependent range
of \(|\bar\theta|\) to be tested by neutron-EDM searches.
Conversely, a neutron-EDM signal attributable predominantly to
\(\bar\theta\), together with a measurement of \(J_{\rm CP}^{\ell}\),
would select branch-labelled intervals in \(m_{\rm lightest}\), which could
then be tested directly by kinematic measurements. 
Likewise, a determination of \(m_{\rm lightest}\) and \(|\bar\theta|\)
would select the compatible leptonic CP branches.

Finite experimental precision and the discrete branch structure can leave
more than one solution. The allowed regions are narrowest where the CP
response varies appreciably with the parameters, while weakly varying or
near-saturated regions can retain broader degeneracies. Increasing
precision in any of the three probes therefore progressively narrows the
allowed branch-labelled regions.

\section{Conclusions}
\label{sec:conclusions}

We have shown that supplementing generalized parity with a sectorial
\(\CPell\) reality condition on the Dirac lepton Yukawa matrices turns the
parity reconstruction of \(V_R^{\ell}\) into a branch-dependent prediction
for observable leptonic CP violation. Under parity alone, \(\dCP\) is an
input to the reconstruction. Leptonic reality instead requires the
right-handed mixing matrix to be rephasing-equivalent to \(V_L^*\), and
compatibility determines the allowed values of \(\dCP\) for each leptonic
branch.

At the Yukawa level, this compatibility problem has a simple exact
structure. For generic three-generation Yukawas, simultaneous parity and
leptonic reality are equivalent to the vanishing of a single
Jarlskog-type CP-odd invariant. The apparent conditions on the
right-handed mixing moduli and CP orientation are therefore not
independent, but are different manifestations of the same physical
compatibility condition. On compatible physical parity solutions, the
PMNS Jarlskog invariant is fixed exactly by the parity-induced shift
\(\Delta J_P\equiv J_R-J_L\) through
\(J_L=-\Delta J_P/2\).

The small Dirac-neutrino masses play a dual role in the parity
reconstruction. Their absolute scale controls the domain in which
physical parity solutions exist, while compressed mixed-sign spectra
contain small signed mass sums that can strongly enhance the leptonic
response. This signed-spectrum enhancement allows values of \(\eps\)
small enough to satisfy the strong-CP constraint to produce order-one
values of the normalized PMNS Jarlskog invariant, including maximal CP
violation.

The compatibility relation derived here is a tree-level result at the
scale where sectorial leptonic reality is imposed. Connecting it
quantitatively to low-energy observables in a complete realization
requires RG evolution of the lepton parameters. The radiative stability
of the reality condition and possible additional contributions to
\(\bar\theta\) must also be examined. The construction is a restricted
leptonic benchmark and does not address the origin of the small
Dirac-neutrino masses.

The discrete leptonic and quark structures enter separately. At fixed
\(\eps\), the leptonic signs control the enhancement pattern through the
signed neutrino masses, while the quark signs determine the
branch-dependent conversion between \(\eps\) and the induced
\(\bar\theta\). Combining the two reconstructions gives a family of
branch-labelled correlations in
\((J_{\rm CP}^{\ell},m_{\rm lightest},|\bar\theta|)\).
Oscillation and absolute neutrino-mass measurements, together with
neutron-EDM limits or a signal attributable predominantly to
\(\bar\theta\), can test this correlation from complementary directions.
A viable realization requires the regions selected by these three probes
to overlap for at least one compatible pairing of leptonic and quark
branches.

\appendix

\section{Explicit response factors and enhanced limits}
\label{app:explicit}
This appendix gives the explicit response factors entering the ordinary
first-order and quasi-degenerate expansions used in the main text. At
either CP-conserving origin,
\[
\delta_0=0,\pi,
\qquad
\zeta\equiv\cos\delta_0=\pm1,
\]
we denote by \(O\equiv V_L(\delta_0)\in SO(3)\) the corresponding real
PMNS representative. In the standard three-angle convention,
{\small
\begin{equation*}
O=
\begin{pmatrix}
c_{12}c_{13}
&
s_{12}c_{13}
&
\zeta s_{13}
\\
-s_{12}c_{23}-\zeta c_{12}s_{23}s_{13}
&
c_{12}c_{23}-\zeta s_{12}s_{23}s_{13}
&
s_{23}c_{13}
\\
s_{12}s_{23}-\zeta c_{12}c_{23}s_{13}
&
-c_{12}s_{23}-\zeta s_{12}c_{23}s_{13}
&
c_{23}c_{13}
\end{pmatrix}.
\end{equation*}
}
 
\subsection{Ordinary first-order response}
\label{app:explicit_firstorder}

We make explicit here how the first-order result of
Sec.~\ref{sec:firstorder} translates into a displacement of the leptonic
CP phase. Near either CP-conserving origin,
\begin{equation*}
\dCP=\delta_0+\eta,
\end{equation*}
with \(\eta=\mathcal O(\eps)\). The PMNS Jarlskog invariant is locally
\begin{equation*}
J_{\rm CP}^{\ell}\simeq \zeta J_{\max}\eta .
\end{equation*}
Using Eq.~\eqref{eq:J_firstorder}, the phase displacement separates into
the three neutrino-pair channels,
\begin{equation}
\eta\simeq\sum_{i<j}\eta^{(ij)},
\label{eq:eta_channel_sum}
\end{equation}
where
\begin{equation}
\eta^{(ij)}\!
=
-\frac{\eps}{2\zeta J_{\max}}
\frac{
n_{ij}(O^T\hat m_\ell O)_{ij}
}{
\hat m_{\nu_i}+\hat m_{\nu_j}
}
=
\frac{\eps m_\tau}
{2(\hat m_{\nu_i}+\hat m_{\nu_j})}
C_{ij}(S_e).
\label{eq:eta_channel_firstorder}
\end{equation}
Here \(n_{ij}\) is defined in Eq.~\eqref{eq:n_def}, and all mixing-angle
dependence is contained in the real matrix \(O=V_L(\delta_0)\). The
dimensionless charged-lepton response is
\begin{equation}
C_{ij}(S_e)
=
\sum_{\alpha=e,\mu,\tau}
s_\alpha\frac{m_\alpha}{m_\tau}\,
C_{ij}^{\alpha},
\qquad
C_{ij}^{\alpha}
=
-\frac{
n_{ij}O_{\alpha i}O_{\alpha j}
}{
\zeta J_{\max}
}.
\label{eq:Cij_full}
\end{equation}
Equation~\eqref{eq:eta_channel_firstorder} displays separately the
signed-neutrino denominator responsible for the enhancement and the
charged-lepton angular response.

Since
\begin{equation*}
\frac{m_\mu}{m_\tau}\simeq5.9\times10^{-2},
\qquad
\frac{m_e}{m_\tau}\simeq2.9\times10^{-4},
\end{equation*}
the charged-lepton response is dominated by the \(\tau\) term,
\begin{equation}
\eta^{(ij)}
\simeq
\frac{\eps\hat m_\tau}
{2(\hat m_{\nu_i}+\hat m_{\nu_j})}\,
C_{ij}^{\tau}.
\label{eq:eta_channel_tau}
\end{equation}
For the oscillation parameters used here, the muon contribution changes
the \(\tau\)-only result by less than about \(10\%\) in the three channels,
while the electron contribution is negligible.

As a representative enhanced case, consider the mixed-sign \(12\) family,
\(S_\nu=(+,-,\pm)\). The corresponding phase displacement is
\begin{equation}
\eta^{(12)}
\simeq
\frac{\eps\hat m_\tau}
{2(m_{\nu_1}-m_{\nu_2})}\,
C_{12}^{\tau},
\label{eq:eta_12_tau_app}
\end{equation}
with the explicit angular coefficient
\begin{equation}
C_{12}^{\tau}
=
s_{23}^{2}
-c_{23}^{2}s_{13}^{2}
-\zeta\,
\frac{c_{12}^{2}-s_{12}^{2}}
{c_{12}s_{12}}\,
s_{13}c_{23}s_{23}.
\label{eq:C12_tau}
\end{equation}
For the central oscillation parameters used here,
\(C_{12}^{\tau}\) remains an order-one factor, approximately
\(0.4\)--\(0.6\) across the two orderings and CP-conserving origins.
Using Eq.~\eqref{eq:solar_difference_identity}, the small signed
denominator in Eq.~\eqref{eq:eta_12_tau_app} is directly controlled by
the solar mass-squared splitting, making the origin of the enhancement
explicit.
 
The compressed atmospheric branch
\(S_\nu=(++-)\) provides a second useful limit.
In this case the enhanced channels are \(13\) and \(23\). Their denominators
become equal at leading order in the compressed limit, so the combined
response is proportional to \(C_{13}^{\tau}+C_{23}^{\tau}\).
Orthogonality gives
\begin{equation*}
C_{12}^{\tau}+C_{13}^{\tau}+C_{23}^{\tau}=0,
\end{equation*}
and hence
\begin{equation}
\eta^{(++-)}
\simeq
\frac{\eps\hat m_\tau m_0}
{\Delta m_{31}^2}\,
C_{12}^{\tau}.
\label{eq:eta_FO_ppm_atm}
\end{equation}
Thus the combined 13 and 23 responses reduce to the coefficient
\(C_{12}^{\tau}\).

\subsection{Quasi-degenerate response}
\label{app:QDexplicit}

In the QD regime, the displacement from a CP-conserving origin may be written as
\begin{equation*}
\dCP=\delta_0+\eta_{\rm QD} .
\end{equation*}
Using the QD result of Sec.~\ref{sec:QD}, the phase displacement separates
into charged-lepton flavor pairs,
\begin{equation}
\eta_{\rm QD}
\simeq
\sum_{\alpha<\beta}\eta_{\rm QD}^{(\alpha\beta)},
\qquad
\eta_{\rm QD}^{(\alpha\beta)}
=
\frac{\kappa_{\alpha\beta}}{2}\,Q_\gamma,
\qquad
\gamma\neq\alpha,\beta ,
\label{eq:eta_QD_C_general}
\end{equation}
where
\begin{equation}
Q_\gamma
\equiv
\Delta_{21}C_{13}^{\gamma}
+
\Delta_{31}C_{12}^{\gamma},
\qquad
\Delta_{i1}\equiv\frac{\Delta m_{i1}^2}{m_0^2}.
\label{eq:Qgamma_def}
\end{equation}
This form follows from the row--column correspondence between the QD
response factors and the ordinary coefficients \(C_{ij}^{\gamma}\) defined in
Eq.~\eqref{eq:Cij_full}. Orthogonality gives
\begin{equation*}
Q_e+Q_\mu+Q_\tau=0 .
\end{equation*}

Consider the mixed QD branch
\((\sigma_e,\sigma_\mu,\sigma_\tau)=(+,+,-)\).
The \(e\mu\) phase pair remains regular, while the \(e\tau\) and
\(\mu\tau\) pairs are enhanced because their QD phase signs are opposite.
In the regime
\(\left|\eps\hat m_\alpha/(2m_0)\right|\ll1\), their leading combined
phase displacement is
\begin{equation}
\eta_{\rm QD}^{(++-)}
\simeq
-\frac{m_0}{\eps}
\left[
\frac{Q_\mu}{\hat m_\tau-\hat m_e}
+
\frac{Q_e}{\hat m_\tau-\hat m_\mu}
\right].
\label{eq:eta_QD_ppm_twochannel}
\end{equation}
Since \(m_e,m_\mu\ll m_\tau\), the two denominators are equal at leading
order. Using \(Q_e+Q_\mu=-Q_\tau\), the interference between the two
enhanced phase pairs therefore reduces to
\begin{equation}
\eta_{\rm QD}^{(++-)}
\simeq
\frac{m_0}{\eps\hat m_\tau}\,Q_\tau
=
\frac{
\Delta m_{21}^2 C_{13}^{\tau}
+
\Delta m_{31}^2 C_{12}^{\tau}
}{
\eps\hat m_\tau m_0
}.
\label{eq:eta_QD_ppm_mass_full}
\end{equation}

Although \(|C_{13}^{\tau}|>|C_{12}^{\tau}|\), the first term is suppressed
by the solar-to-atmospheric mass-squared ratio and changes the leading
result by only about \(10\%\) for the numerical inputs used here. 
The atmospheric contribution therefore gives
\begin{equation}
\eta_{\rm QD}^{(++-)}
\simeq
\frac{\Delta m_{31}^2}
{\eps\hat m_\tau m_0}\,
C_{12}^{\tau}.
\label{eq:eta_QD_ppm_mass}
\end{equation}
Thus the interference of the enhanced \(e\tau\) and \(\mu\tau\) phase
pairs selects the same angular coefficient \(C_{12}^{\tau}\) that appears
in the ordinary atmospheric response. For the branch used in the main
text, \(\delta_0=\pi\) and hence \(\zeta=-1\), so
\(J_{\rm CP}^{\ell}/J_{\max}\simeq-\eta_{\rm QD}\), reproducing the QD
asymptotic form in Eq.~\eqref{eq:atm_two_asymptotics}.

\section{Numerical inputs and branch conventions}
\label{app:numerical_details}

This appendix collects the numerical inputs, branch identifications, and
normalization conventions used in Sec.~\ref{sec:phenomenology}.

\paragraph{Leptonic inputs.}
We use the ordering-dependent NuFIT~v6.1 best-fit values including
atmospheric data \cite{Esteban:2024eli,NuFIT:v61},
\begin{align*}
(\theta_{12},\theta_{23},\theta_{13})_{\rm NO}
&=
(33.76^\circ,43.29^\circ,8.62^\circ),
\\
(\theta_{12},\theta_{23},\theta_{13})_{\rm IO}
&=
(33.76^\circ,47.90^\circ,8.65^\circ),
\end{align*}
with
\begin{equation*}
\begin{aligned}
\Delta m_{21}^2
&=
7.537\times10^{-5}\ {\rm eV}^2,
\\
\Delta m_{31}^2
&=
+2.511\times10^{-3}\ {\rm eV}^2,
\qquad {\rm NO},
\\
\Delta m_{32}^2
&=
-2.483\times10^{-3}\ {\rm eV}^2,
\qquad {\rm IO}.
\end{aligned}
\end{equation*}
Thus \(m_{\rm lightest}=m_{\nu_1}\) in NO and
\(m_{\rm lightest}=m_{\nu_3}\) in IO. We use the charged-lepton masses
\(m_e=0.5110\,{\rm MeV}\),
\(m_\mu=105.658\,{\rm MeV}\), and
\(m_\tau=1776.86\,{\rm MeV}\).
The quoted digits specify the central inputs used to reproduce the
numerical solutions. No leptonic RG evolution is applied to these inputs.

\paragraph{Compatibility condition.}
The Yukawa and mass commutators obey the exact relation
\begin{equation*}
[M_\nu,M_\ell]
=
v^2\cos2\beta\,[Y_1^\ell,Y_2^\ell].
\end{equation*}
Within the physical hierarchy reconstruction used numerically,
\[
M_\ell=\hat m_\ell,
\qquad
M_\nu=V_LX^\dagger V_L^\dagger ,
\]
so, for \(\cos2\beta\neq0\), the compatibility condition
\(\mathcal I_Y=0\) becomes
\begin{equation*}
\Tr\!\left([X^\dagger,H]^3\right)=0,
\end{equation*}
as stated in Eq.~\eqref{eq:IY_X_zero}.

For physical solutions, unitarity together with
Eq.~\eqref{eq:master} implies
\begin{equation*}
X-X^\dagger=i\eps H.
\end{equation*}
Hence
\[
[X^\dagger,H]
=
\left[\frac{X+X^\dagger}{2},H\right],
\]
which is anti-Hermitian. Its cubic trace is therefore purely imaginary,
so the compatibility equation imposes one real condition.

\paragraph{Leptonic branch counting and symmetries.}
For each mass ordering there are \(8\times8\) nominal parity assignments
of \((S_e,S_\nu)\). The exact identification
\[
(S_e,S_\nu)\sim(-S_e,-S_\nu)
\]
reduces these to \(32\) parity-sign classes. Compatibility introduces two
CP-conserving origins, \(\delta_0=0,\pi\), for each class. Continuing both
origins to nonzero \(\eps\) therefore gives \(64\) seed-labelled
compatibility branches.

For the hierarchy reconstruction used numerically, the sign of \(\eps\)
is redundant once all charged-lepton sign branches are included. Indeed,
the transformation
\[
\eps\longrightarrow-\eps,
\qquad
S_e\longrightarrow-S_e
\]
leaves the reconstruction equations unchanged and changes \(V_R^\ell\)
only by an overall sign. Since all \(S_e\) branches are included, the
complete branch set may therefore be represented as a function of
\(|\eps|\) without loss of solutions.

Complex conjugation gives a separate relation. It maps a compatible
solution at \(\eps\) to its CP-conjugate solution at \(-\eps\), with
\[
J_{\rm CP}^{\ell}\longrightarrow-J_{\rm CP}^{\ell}.
\]
Combining this relation with the sign redundancy above maps the conjugate
solution back to positive \(|\eps|\) on the globally sign-reversed
charged-lepton branch. Consequently, the complete solution set at fixed
\(|\eps|\) is symmetric under
\[
J_{\rm CP}^{\ell}\longrightarrow-J_{\rm CP}^{\ell}.
\]

In the \(m_e\to0\) limit the electron sign becomes irrelevant and the
\(64\) seed-labelled branches collapse into \(32\) effective families,
with \(16\) sign patterns associated with each CP-conserving origin.
Finite-\(m_e\) effects weakly split some otherwise coincident curves.

\paragraph{Quark normalization.}
For the conversion between \(\eps\) and \(\bar\theta\), we take
\(\mu_P=10\,{\rm TeV}\) as a representative common scale and evaluate
Eq.~\eqref{eq:Cq_branch} using the central \(\overline{\rm MS}\) SM
running quark Yukawas and CKM parameters obtained from the 2024 PDG inputs
of Ref.~\cite{Antusch:2025fpm}. Across the inequivalent quark sign classes
this gives
\[
12.88\lesssim |C_q^{(S)}|\lesssim50.25,
\]
which yields the reference values quoted in
Sec.~\ref{subsec:numerical_inputs}. Varying the common scale between
\(1\) and \(100\,{\rm TeV}\) changes these reference values by less than
about \(10\%\). Hadronic and quark-input uncertainties are not propagated.

\section{PMNS-phase representation of the branch structure}
\label{app:delta_branches}

\begin{figure}[t]
\centering
\makebox[\linewidth][c]{%
  \hspace*{2pt}%
  \includegraphics[width=1.02\linewidth]{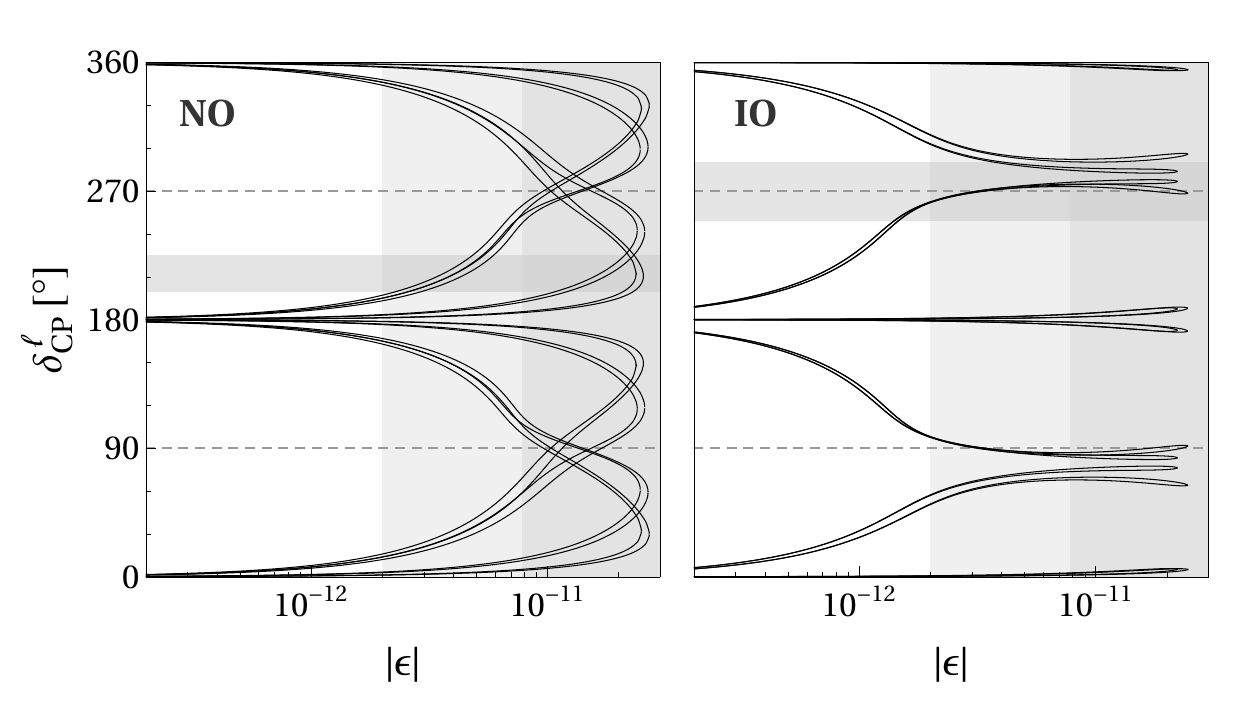}%
  \hspace*{4pt}%
}
\caption{
PMNS-phase representation of the nonlinear fixed-mass solutions at
\(m_{\rm lightest}=0.01\,{\rm eV}\), corresponding to
Fig.~\ref{fig:J_response_light}. We show \(\dCP\) in degrees as a function
of \(|\eps|\), with all leptonic compatibility branches included.
The vertical shading follows the strong-CP convention of the invariant
plot, while the darker horizontal bands mark
\(\dCP\in[200^\circ,225^\circ]\) for NO and
\(\dCP\in[250^\circ,290^\circ]\) for IO.
}
\label{fig:delta_response_001}
\end{figure}

\begin{figure}[H]
\centering
\makebox[\linewidth][c]{%
  \hspace*{2pt}%
  \includegraphics[width=1.02\linewidth]{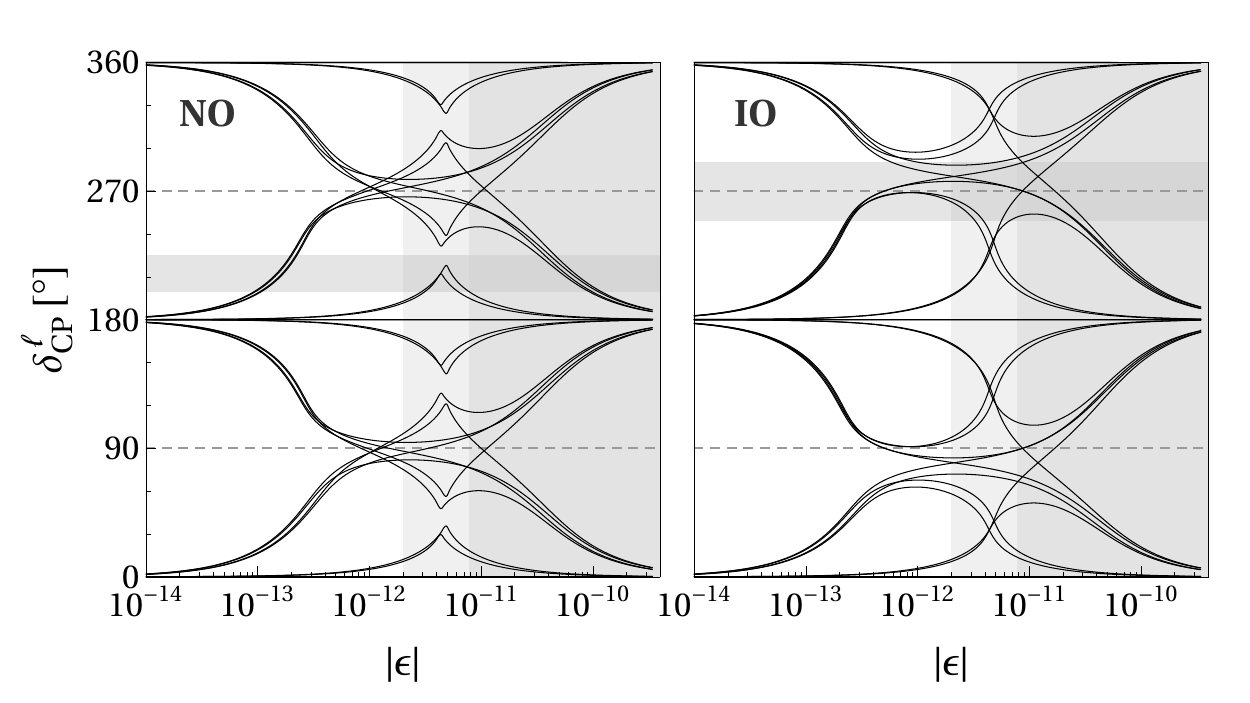}%
  \hspace*{4pt}%
}
\caption{
PMNS-phase representation of the nonlinear fixed-mass solutions at the
diagnostic benchmark \(m_{\rm lightest}=0.3\,{\rm eV}\), corresponding to
Fig.~\ref{fig:J_response_QD}. We show \(\dCP\) in degrees as a function of
\(|\eps|\), with all leptonic compatibility branches included. The vertical
and horizontal shading follows the same conventions as
Fig.~\ref{fig:delta_response_001}. In the compressed spectrum, the direct
phase representation resolves the pronounced branch evolution near
\(\dCP=90^\circ\) and \(270^\circ\) that is compressed in the invariant
plot by the saturation of
\(J_{\rm CP}^{\ell}/J_{\max}=\sin\dCP\).
}
\label{fig:delta_response_03}
\end{figure}

 The main text presents the numerical solutions in terms of the invariant
ratio \(J_{\rm CP}^{\ell}/J_{\max}\). Since
\(J_{\rm CP}^{\ell}/J_{\max}=\sin\dCP\), this representation is not
one-to-one: the two CP-conserving origins, \(\dCP=0\) and \(\pi\), both
map to zero, while the phase evolution is compressed near maximal CP
violation, where the sine saturates.
Figures~\ref{fig:delta_response_001} and
\ref{fig:delta_response_03} show the same fixed-mass solutions as
Figs.~\ref{fig:J_response_light} and \ref{fig:J_response_QD},
respectively, with the same horizontal coordinate and shading but with
\(\dCP\) on the vertical axis. The phase is shown in degrees for direct
comparison with the representative intervals used in
Sec.~\ref{subsec:compatibility_bands}.

\newpage

  \FloatBarrier

\bibliographystyle{apsrev4-2}
\bibliography{refs}
 
\end{document}